\documentclass[11pt]{article}
\usepackage{moriond,epsfig}

\def\Journal#1#2#3#4{{#1} {\bf #2}, #3 (#4)}

\def\A&A{\em Astron. Astroph.}
\def\APP{\em Astropart. Phys.}

\def\APL{\em Appl. Phys. Lett.}
\def\JAP{\em J. Appl. Phys.}

\def\NIMA{{\em Nucl. Instrum. Methods} A}

\def\PLB{{\em Phys. Lett.}  B}

\def\PRL{\em Phys. Rev. Lett.}
\def\PRD{{\em Phys. Rev.} D}

\def\be{\begin{equation}}
\def\ee{\end{equation}}
\def\bea{\begin{eqnarray}}
\def\eea{\end{eqnarray}}

\def\etal{{\it et al.}}

\begin{document}
\vspace*{4cm}
\title{Dark Matter Search in the EDELWEISS experiment}

\author{ A. JUILLARD$^{1}$, A. BENOIT$^{2}$, L. BERGE$^{3}$, A. BONNEVAUX$^{4}$, R. BOUVIER$^{4}$,\\ A. BRONIATOWSKI$^{3}$,
B. CHAMBON$^{4}$, M. CHAPELLIER$^{5}$, G. CHARDIN$^{1}$, P. CHARVIN$^{1,6}$,\\ P. CLUZEL$^{4}$, M. DE JESUS$^{4}$,
P. DI STEFANO$^{1}$, D. DRAIN$^{4}$, L. DUMOULIN$^{3}$, J. GASCON$^{4}$,\\ G. GERBIER$^{1}$, C. GOLDBACH$^{7}$,
M. GOYOT$^{4}$, M. GROS$^{1}$,J.P. HADJOUT$^{4}$,  A. DE LESQUEN$^{1}$,\\ M. LOIDL$^{1}$, D. LOISEAU$^{1}$,
J. MALLET$^{1}$, S. MARNIEROS$^{3}$, O. MARTINEAU$^{4}$,\\ N. MIRABOLFATHI$^{3}$, L. MOSCA$^{1}$,
X-F. NAVICK$^{1}$, G. NOLLEZ$^{7}$, P. PARI$^{5}$, C. PASTOR$^{4}$,\\ E. SIMON$^{4}$, M. STERN$^{4}$, L. VAGNERON$^{4}$}

\address{$^{1}$CEA, Centre d'Etudes Nucl\'eaires de Saclay, DSM/DAPNIA, F-91191 Gif-sur-Yvette, France\\
$^{2}$Centre de Recherche sur les Tr\`es Basses Temp\'eratures, BP 166, 38042 Grenoble, France\\
$^{3}$CSNSM, IN2P3-CNRS, Univ. Paris XI, bat. 108, F-91405 Orsay Cedex, France\\
$^{4}$IPN Lyon and UCBL, IN2P3-CNRS, 43 Bd. 11 novembre 1918, F-69622 Villeurbanne Cedex, France\\
$^{5}$CEA, Centre d'Etudes Nucl\'eaires de Saclay, DSM/DRECAM, F-91191 Gif-sur-Yvette Cedex, France\\
$^{6}$Laboratoire Souterrain de Modane, CEA-CNRS, 90 rue Polset, F-73500 Modane, France\\
$^{7}$Institut d'Astrophysique de Paris, INSU-CNRS, 98 bis Bd. Arago, F-75014 Paris, France\\}

\maketitle\abstracts{
The EDELWEISS Dark Matter Search uses low-temperature Ge detectors with heat and ionisation read-out to identify nuclear recoils induced by elastic collisions with WIMPs from the galactic halo. Preliminary results obtained with 320g bolometers  are described. After a few weeks of data taking, data accumulated with one of these detectors already allow to reach the upper part of the DAMA region. Prospects for the present run and the second stage of the experiment, EDELWEISS-II, using an innovative reversed cryostat allowing data taking with 100 detectors, are briefly described.}

\section{Introduction}
The EDELWEISS experiment is a WIMP direct detection experiment set in the Fr\'ejus Underground Laboratory, adjacent to a highway tunnel connecting France and Italy under the Alps. The rock overburden reduces the muon background flux by a factor $\approx2\times10^6$ while the neutron flux, originating mainly from the surrounding rock, is $\approx4\times10^{-6}$ neutron$/cm^2/s$ ~\cite{Chazal}. 
The experimental setup of the EDELWEISS experiment has been described elsewhere~\cite{bellefon}. EDELWEISS has developed cryogenic germanium detectors with simultaneous measurement of charge and phonon signals. A detailed description of these detectors can be found in L'Hote \etal\ and Navick \etal\cite{lhote,navick}. A first motivation for the development of cryogenic detectors lies in the possibility to lower the energy threshold compared to classical detectors~\cite{crst1}. A second motivation comes from the fact that present experiments appear to be mostly limited by the radioactive background rate of the detectors~\cite{baudis,dama,saclaynai} and by the systematic uncertainties of the rejection scheme using pulse shape discrimination (PSD) techniques in NaI crystals~\cite{saclaynai,ukdmc2}. The simultaneous measurement of the charge and phonon signals using cryogenic detectors~\cite{shutt92}, or light and phonon signals \cite{crst1,crst2} allows a much more reliable discrimination between the main source of radioactive background, producing electron recoils, and the nuclear recoils expected from WIMP interactions. The discrimination performances of the first 70 g EDELWEISS detectors have been analyzed in Di Stefano \etal\cite{stefano} and the limitations brought by surface electron or nuclear interactions have been discussed in Benoit \etal\cite{benoit}. In the following, we discuss the effective rejection factor obtained with a new 320g bolometer with guard ring and aluminum electrodes.

\section{The 320g bolometer}

\begin{figure}
\begin{center}
\psfig{figure=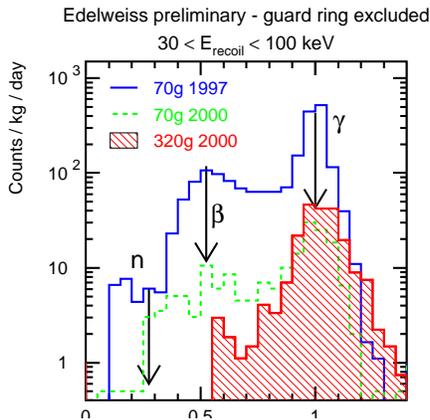,height=2.4in}
\end{center}
\caption{ Ratio of the ionization yield to the recoil energy (quenching factor) for events with recoil energies between 30 and 100 keV. The electron recoil yield has been normalised to 1 using $\gamma$-ray calibration sources and is expected to be approximately 0.3 and 0.5 for nuclear and surface electron recoils, respectively. The data set, normalised to 1 kg$\cdot$day, are: (line) 1.8 kg$\cdot$day exposure for a 70~g detector in the 1997 configuration; (dashed) 2.0 kg$\cdot$day exposure for a 70~g detector in the 2000 configuration and (hatched histogram) 3.1 kg$\cdot$day exposure in the center fiducial region of a 320~g detector with guard ring.
\label{fig:fig1}}
\end{figure}

This 320 g germanium bolometer has been described in Navick \etal\cite{navick}. It is equipped with aluminum electrodes, used as a possible solution to reduce charge collection problems in surface events with respect to implanted electrodes. This detector benefits from a guard ring identifying signals occurring near the edges of the detector. At present, two 320g Ge bolometers have been tested. The complete set of three 320 g bolometers should be installed at the end of the present run. 

On the lower 320 g Ge detector, both central and guard ring electrodes are polarized at 6.3 volts. A linear cross-talk between the two channels (around 10 percent) has been observed, but this relatively limited effect can be accounted for by the simultaneous analysis of signals recorded on both electrodes. A fiducial volume, estimated to 50 percent of the detector volume from neutron calibrations, is defined by selecting events with notable amplitude on the central electrode only. The temperature increase due to an interaction is measured through a DC-polarised NTD germanium sensor, with a resistance around 4M$\Omega$. The temperature is stabilized within  $10\mu$K around 27 mK.
The ionization channel resolutions are 2 keV FWHM baseline and 3keV @ 122keV for the central zone, with a threshold of  $\approx$6 keV e.e., mostly limited by microphonics, and 1.5 keV baseline and 2keV @ 122keV for the guard ring, with a threshold of  $\approx$5 keV e.e. The heat channel exhibits a resolution of 1.5 keV baseline and 3keV @ 122keV.
 The raw data trigger rate, prior to software analysis, is of the order of a few events per minute.

\section{Preliminary results}

We present here the preliminary results obtained in a physics data taking of 6.3 kg $\times$  day, and a fiducial volume exposure of 3.1 kg $\times$ day. A comparison of the performances of the present 320g detectors with the previous 70g detectors has been represented in Fig.1. In this figure, the histogram of the ionization amplitude/recoil energy ratio (the quenching factor), normalized to 1 for electron recoils, has been plotted. It can readily be seen that the performance increase is important. In particular, by restricting the data to the central region only, no events are observed in the [30, 100] keV recoil energy interval over the 2$\sigma$   recoil zone (Fig.2), which corresponds to a nuclear recoil acceptance of $\approx$ 0.95 for the central part of the detector.  A paraffin shield of 30cm surrounding the cryostat has been used to reduce the fast neutron flux from the environment by more than two orders of magnitude. 

\begin{figure}
\begin{center}
\psfig{figure=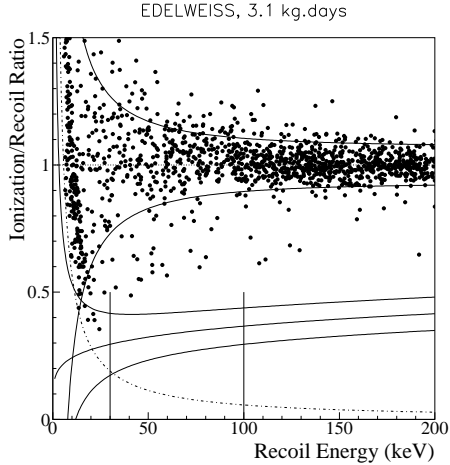,height=2.5in}
\end{center}
\caption{ Scatter diagram of the quenching factor vs. recoil energy for the fiducial region of the 320g detector with guard ring (3.1kg$\cdot$day exposure). No events are observed over the $\pm$2$\sigma$ recoil zone 
\label{fig:fig2}}
\end{figure}

A conservative exclusion contour in the WIMP mass vs. cross-section scatter diagram has been derived from these data, and is plotted in Fig. 3. It can be seen that the upper part of the region associated with the DAMA region\cite{dama} is excluded. From the absence of any recorded event in the [30, 100] keV recoil energy, a significant sensitivity increase over the next few months can reasonably be expected. It is important to note that this limit is obtained without any neutron background subtraction.

\begin{figure}
\begin{center}
\psfig{figure=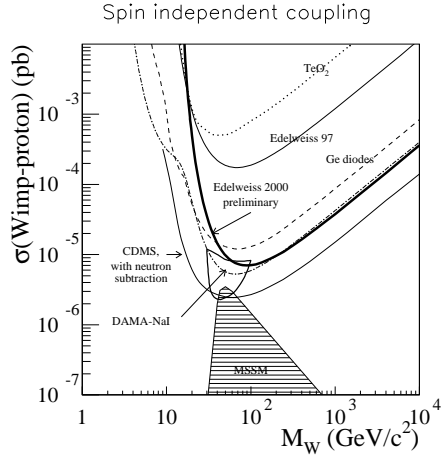,height=2.5in}
\end{center}
\caption{ Scalar coupling WIMP-nucleon cross-section upper limits obtained by the EDELWEISS experiment and compared to the main direct detection experiments. The regions above the curves are excluded at 90\% CL. Also represented are the allowed MSSM models compatible with accelerator data, the region selected by the DAMA experiment, and the sensitivities of some of the main direct detection experiments.
\label{fig:fig3}}
\end{figure}

\section{Perspectives and conclusion}

With an effective mass of 160 gram, $\approx$ 600 times smaller than the DAMA detector, and with an exposure $\approx$ 10 000 times smaller, the present detector already reaches the DAMA region~\cite{dama}. In a forthcoming run, we will install three 320 g Ge detectors similar to the present one. Data taking realized during the present Ò1 kgÓ stage are expected to last until the end of 2001 and should reach the sensitivity required by the most favorable SUSY models. Improvements should also allow to lower the energy threshold by a factor $\approx$ 2.
To explore more deeply the parameter space of SUSY models, it is clear, from the absence of any candidate event after $\approx$ 6 weeks of data taking, that a larger number of detectors will be required. Therefore, we are actively preparing a second stage of the experiment, EDELWEISS-II, which will use a 100-liter dilution fridge of novel geometry and presently built in CRTBT Grenoble. This large cryostat, able to accommodate 100 detectors and their electronics, is expected to be installed in the Frejus Underground Laboratory in 2002. This second stage is expected to provide an increase in sensitivity by more than two orders of magnitude over the best present performances.

\section*{Acknowledgments}
We thank the technical staff of the LSM and of the participating laboratories for their invaluable help. This work has been partially funded by the EEC-Network program under contract ERBFMRXCT980167.


\section*{References}
\begin{thebibliography}{99}
\bibitem{Chazal} V. Chazal \etal, \Journal{\APP}{9} {163} {1998}.
\bibitem{bellefon} A. de Bellefon \etal, \Journal{\APP}{6} {35} {1996}.
\bibitem{lhote} D. L'Hote \etal, \Journal{\JAP}{87} {1507} {2000}.
\bibitem{navick} X-F. Navick \etal, in {\em Proc. 8th Int. Workshop on Low Temperature Detectors} (LTD8), Dalfsen, Netherlands, August 15-20, 1999, \Journal{\NIMA}{444} {361} {2000}.
\bibitem{crst1} M. Bravin \etal, \Journal{\APP}{12} {107} {1999}.
\bibitem{baudis} L. Baudis \etal, \Journal{\PRD}{59} {022001} {1999}.
\bibitem{dama} R.Bernabei \etal, \Journal{\PLB}{389} {757} {1996}; \Journal{\PLB}{424} {195} {1998} ; \Journal{\PLB}{450} {448} {1999}.
\bibitem{saclaynai} G. Gerbier \etal, \Journal{\APP}{11} {287} {1999}. 
\bibitem{ukdmc2} N.J.T. Smith, J.D. Lewin and P.F. Smith, \Journal{\PLB}{485} {1} {2000}.
\bibitem{shutt92} T. Shutt \etal, \Journal{\PRL}{69} {3425} {1992}.
\bibitem{crst2} P. Meunier \etal, \Journal{\APL}{75} {1335} {1999}.
\bibitem{stefano} P. Di Stefano \etal, \Journal{\APP}{14} {329} {2001}.
\bibitem{benoit} A. Benoit \etal, \Journal{\PLB}{479} {8} {2000}.


\end{thebibliography}
\end{document}